\documentclass[]{elsart}

\usepackage{icarus}

\usepackage{natbib}

\bibpunct{(}{)}{;}{a}{,}{,}

\usepackage{graphicx}

\usepackage{amssymb}

\begin{document}

\begin{frontmatter}


\title{Scattering matrices and expansion coefficients \\
       of Martian analogue palagonite particles}


\author[label1,label2]{E.C. Laan,}
\author[label1]{H. Volten,}\footnote{Currently working at the 
                            National Institute for Public Health 
                            and the Environment (RIVM),
                            P.O.~Box 1, 3720 BA Bilthoven, The Netherlands}
\author[label4,label5]{D.~M. Stam,}
\author[label6]{O. Mu\~{n}oz,}
\author[label1]{J.~W. Hovenier,}
\author[label7]{T. L. Roush}

\address[label1]{Astronomical Institute 'Anton Pannekoek',
                 University of Amsterdam, \\ Kruislaan 403, 1098 SJ Amsterdam,
                 The Netherlands}
\address[label2]{TNO Science \& Industry, Stieltjesweg 1, 2600 AD, Delft, The Netherlands}
\address[label4]{SRON Netherlands Institute for Space Research, Sorbonnelaan 2,
                 3584 CA Utrecht, The Netherlands}
\address[label5]{DEOS, Aerospace Engineering, Technical University Delft,
                 Kluyverweg 1, 2629 HS, Delft, The Netherlands}
\address[label6]{Instituto de Astrof\'{i}sica de Andaluc\'{i}a (CSIC),
        	 C/ Camino Bajo de Hu\'{e}tor, 50, 18008, Granada, Spain}
\address[label7]{NASA Ames Research Center, MS 245-3, 
                 Moffett Field, CA 94035-1000, USA}

\end{frontmatter}

                                                                                
                                                                                
\begin{flushleft}
\vspace{1cm}
Number of pages: \pageref{lastpage} \\
Number of figures: \ref{lastfig}\\
Number of tables: 1 \\
\end{flushleft}

                                                                                
\begin{pagetwo}{Scattering matrices of Martian analogue palagonite}
                                                                                
E.C. Laan \\
TNO Science \& Industry\\
Stieltjesweg 1 \\
2600 AD Delft, The Netherlands \\

Email: eriklaan@dds.nl \\
Phone: +31 6 47974621 \\
Fax: +31 15 2692111 \\ 
                                                                                
\end{pagetwo}

\begin{abstract}

{\em
We present measurements of ratios of elements of the scattering matrix of Martian 
analogue palagonite particles for scattering angles
ranging from 3$^\circ$ to 174$^\circ$ and a wavelength of 632.8~nm.
To facilitate the use of these measurements in radiative transfer calculations
we have devised a method that enables us to obtain, from these measurements, 
a normalized synthetic scattering matrix covering the complete scattering angle range 
from 0$^\circ$ to 180$^\circ$.
Our method is based on employing the coefficients of the expansions of 
scattering matrix elements into generalized spherical functions. 
The synthetic scattering matrix elements and/or the expansion coefficients 
obtained in this way, can 
be used to include multiple scattering by these irregularly shaped particles 
in (polarized) radiative transfer calculations, such as calculations of 
sunlight that is scattered in the dusty Martian atmosphere.
}

\end{abstract}


\section{Introduction}
\label{section_introduction}

Dust from the Martian surface is regularly swept up by winds to form
local, regional, or sometimes even planet-wide dust storms.
The airborne dust particles scatter and absorb solar radiation 
and are therefore very important for the thermal structure of 
the thin Martian atmosphere and for the temperature of the Martian 
surface \citep[see e.g.][and references therein]{2006GeoRL..3302203B,
2005Icar..175...23G,2004Icar..167..148S,1972JAtS...29..400G}. 
The interaction between radiation and the dust particles thus 
has to be taken into account when studying, for
example, the local and global climate on Mars. 
In particular, one has to account for the spatial distribution of the
dust particles, their number density, and their optical
properties.
For a given wavelength, the optical properties of the dust particles
depend on their composition, sizes and shapes. 

Despite the important role of dust particles in the Martian atmosphere,
surprisingly little is known about their optical properties
\citep[for an overview, see][]{2005AdSpR..35...21K}. 
Consequently, in radiative transfer calculations that are used to 
interpret space or ground-based observations of Mars, various 
assumptions are made regarding the dust optical properties.  
In particular, it is common to assume homogeneous, spherical or 
spheroidal dust particles \citep{1995JGR...100.5235P,1999JGR...104.8987T},
although dust particles on Earth are known to be irregularly shaped.
The optical properties of homogeneous, spherical particles can 
straightforwardly be calculated using Lorenz-Mie theory
\citep[][]{1957lssp.book.....V,1984A&A...131..237D}, and those
of homogeneous, spheroidal particles using
e.g. the so-called T-matrix method \citep[][]{1994OptCo.109...16M,2006JGRD..11111208D}.
The optical properties of spherical particles can, however, differ 
significantly from those of irregularly shaped particles, even if 
their composition and/or size distribution is the same.
Therefore, assuming spherical instead of 
irregularly shaped particles in radiative transfer calculations
that are used for example to analyze observations, can lead to 
significant errors in retrieved atmospheric parameters, such as the
dust optical thickness and/or dust particle size distributions
\citep[for a discussion on such errors, 
see e.g.][]{2003SoSyR..37...87D,2002SoSyR..36..367D}. 

For irregularly shaped particles, which are very common in nature,
the scattering matrix elements can in principle be calculated with
numerical methods such as those based on the so--called Discrete Dipole
Approximation (DDA) method \citep[see e.g.][]{1994OSAJ...11.1491D}.
However, the vast amounts of computing time that such
numerical calculations require make it at least very impractical
to calculate complete scattering matrices for a sample of
particles with various (irregular) shapes and various sizes,
in particular if the particles are large compared to the 
wavelength of the scattered light.
Alternatively, employing geometrical optics for unrealistically 
spiky particles combined with an imaginary part of the refractive 
index that is rather small compared to the typical values 
used in the literature,
appears to be useful to reproduce the scattering behavior of 
irregularly shaped mineral particles 
\citep[see][]{2003JGRD.108aAAC12N}.
As suggested by e.g. \citet{1999JGR...104.8987T,2003SoSyR..37...87D,2003JGRE.108i....1W}, 
a more practical method to obtain elements of the scattering matrix
for an ensemble of irregularly shaped particles is to measure the
elements in a laboratory.
Note that measured scattering matrix elements are also essential
for validating numerical methods and approximations.

In this article, we present measurements of ratios of elements of the scattering 
matrix of irregularly shaped, randomly oriented 
Martian analogue palagonite particles, described by \citet{1997JGR...10213341B}, 
as functions of the scattering angle.
The material palagonite is believed to be a reasonable, but not perfect, analogue for the Martian surface and atmospheric dust particles.
Terrestrial palagonite particles (i.e. terrestrial weathering
products of basaltic ash or glass) have been put forward as 
Martian dust analogues \citep[][]{1981LPI....12..271E,1995JGR...100.5309R} 
because of spectral similarities observed with visible and near-infrared 
spectroscopic observations using both Earth-based telescopes and 
several Mars orbiting spacecraft  
\citep[][]{1985AdSpR...5Q..59S,1992mars.book..557S}. 

The measurements have been performed using a HeNe laser, which has
a wavelength of 632.8~nm, and for
scattering angles, $\Theta$,  ranging from 3$^\circ$ (near-forward scattering) 
to 174$^\circ$ (near-backward scattering). 
Other examples of such measurements for irregularly shaped mineral particles obtained with the 
same experimental set-up have been reported by e.g.
\citet{2005JQSRT..90..191V} and references therein. 
Since we have measured ratios of all (non-zero) elements of the
scattering matrix as functions of the scattering angle,
our results can be used for radiative transfer
calculations that include multiple scattering and polarization.
As described by e.g. \citet{1998GeoRL..25..135L}, ignoring polarization, i.e. using only scattering matrix element (1,1) (the so-called phase function), in
multiple scattering calculations, induces errors in calculated fluxes. 
The use of only the phase function should be limited to single scattering calculations for unpolarized incident light. 

A practical limitation of our experimental method is that we 
cannot measure close ($< 6^\circ$) to the exact backscattering direction 
($\Theta=180^\circ$), because there our detector would 
interfere with the incoming beam of light, 
nor close ($< 3^\circ$) to the exact forward scattering 
direction ($\Theta=0^\circ$),
because there our detector would intercept the unscattered 
part of the incident beam.
These two scattering directions are, however, important for
radiative transfer applications. This holds in particular for the near-forward
scattering direction,
since a significant fraction of the light that is incident 
on a particle is generally scattered in the near-forward direction. 
A solution to the lack of measurements in the near-forward and 
near-backward scattering directions is to add artificial 
data points. For the near-forward scattering direction, where a 
strong peak in the phase function is expected, one can add
artificial data points calculated using e.g. Lorenz-Mie calculations.
This has been done before, e.g. by 
\citet{2007JQSRT.103...27K,2003JQSRT..79..903L,2004GeoRL..3104113V,2005JGRD..11010S02H}.
In this article, we use a method similar to that of \cite{2003JQSRT..79..903L},
that was also used by e.g. \cite{2007JGRD..11213215M}, but we extend 
this by using expansion coefficients
which result from a Singular Value Decomposition (SVD) 
fit to the measurements with generalized 
spherical functions \citep[][]{1963Gelfand,1983A&A...128....1H,1984A&A...131..237D,
2004Hovenier}.
With the added artificial data points and the expansion coefficients, we construct a so--called synthetic 
scattering matrix, which is normalized so that the average of the synthetic phase function over all directions equals unity and covers the whole scattering angle range 
(i.e. from 0$^\circ$ to 180$^\circ$). 

Tables of the measurements, the synthetic
scattering matrix elements and the expansion coefficients will be
available from the Amsterdam 
Light Scattering Database\footnote{The Amsterdam Light Scattering
Database is located at: http://www.astro.uva.nl/scatter}. See  
\citet{2005JQSRT..90..191V,2006JQSRT.100..437V} for a description of this database.

The structure of this article is as follows.
In Sect.~\ref{section_palagonite}, we describe the microphysical 
properties of our  Martian analogue palagonite dust particles.
In Sect.~\ref{section_scatteringmatrix}, we define 
the scattering matrix, describe the experimental set-up,
and present the measurements and an auxiliary scattering matrix.
In Sect.~\ref{section_expansioncoefs}, we introduce the
expansion coefficients, describe the Singular Value Decomposition
fitting method, and present the derived expansion coefficients and 
synthetic scattering matrix. 
In Sect.~\ref{section_summary}, finally, we summarize and discuss our results. 

\section{Martian analogue palagonite particles}
\label{section_palagonite}

Palagonite is a fine--grained weathering product of basaltic glass. 
At visible wavelengths it has a refractive index, $m$, typical for 
silicate materials, i.e. Re($m$) is about 1.5 and Im($m$) is 
in the range $10^{-3}$ to $10^{-4}$ \citep[][]{1995JGR...100.5251C}. 
Palagonite contains a considerable fraction 
(about 10\% by mass) of iron (III) oxide (Fe$_{2}$O$_{3}$)
\citep[][]{1995JGR...100.5309R}, which gives Mars its reddish color. 

The sample in this study is sample 91-16 that is 
described in detail by \citet{1997JGR...10213341B}. 
Note that there is another Martian analogue palagonite sample 
described in the literature, namely sample 91-1
\citep[see][]{1995JGR...100.5309R}. 
Palagonite sample 91-1 appears to contain more sodium than sample 91-16 
because of evaporation and deposition of salt due to the proximity of its 
retrieval site to the Pacific Ocean. 
Palagonite sample 91-16 was retrieved at the top of Hawaii's 
Mauna Kea volcano, about 4~km above sea level, where it was formed
in a semi--arid environment likely associated with ephemeral 
melting water from ice. Hence, sample 91-16 is
considered to be the better alternative for Martian dust
of the two Martian palagonite analogues. 
 
\begin{figure}
\vspace*{2.0cm}
\begin{center}
\includegraphics[width=15.0cm,angle=0]{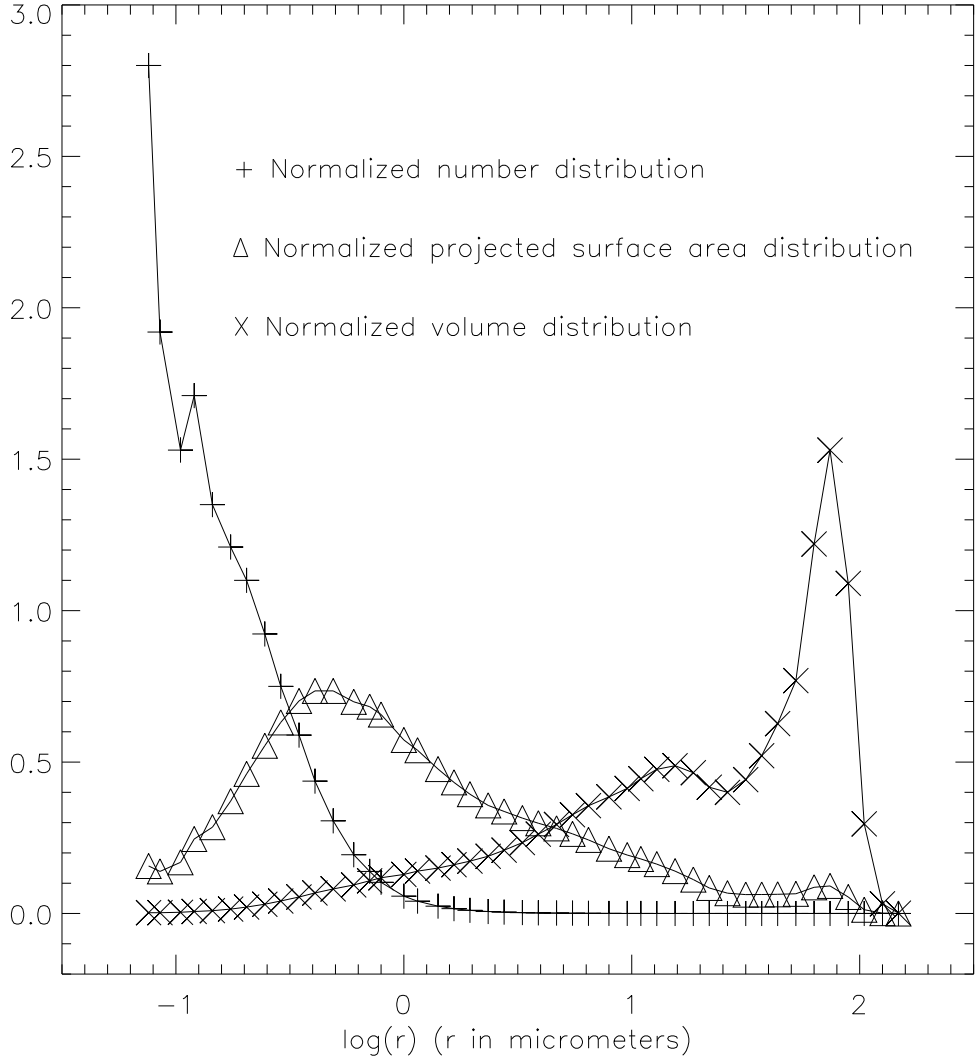}
\end{center}
\vspace*{-0.0cm}
\caption{Measured size distributions of the Martian analogue
         palagonite particles.
         The lines refer to the normalized number distribution 
	 (plusses), the normalized projected-surface-area distribution
	 (triangles), and the normalized volume distribution (crosses)
         as functions of $\log r$, with $r$ the radius (in microns) of a 
         projected-surface-area-equivalent sphere  
         \citep[see][for the definitions of the 
         various distributions]{2005JQSRT..90..191V}.
         Approximating the normalized number distribution with 
         a log normal distribution function yields 
         $r_{\rm eff}=4.46~\mu$m and $v_{\rm eff}=7.29$.} 
\label{fig_size}
\end{figure}

Before using sample 91-16 in our light scattering experiment, 
we removed the millimeter-sized particles by using a sieve with
a 200-$\mu$m grid width, to avoid clogging
the aerosol generator.
Figure~\ref{fig_sem} shows an image of the palagonite particles 
obtained with a scanning electron microscope (SEM).
This image clearly shows the irregular shapes of the palagonite 
particles.
It should be noted that SEM images are not necessarily representative 
of the size distribution of the particles.
The normalized projected-surface-area distribution of the dust 
particles was measured by using a laser particle sizer
that is based on diffraction without
making assumptions about the refractive indices of the materials
of the particles \citep[][]{Konert1997}.
From the projected-surface-area distribution, we derive the 
number distribution and the volume distribution of the particles
because these distributions are often required for numerical
applications.
Figure~\ref{fig_size} shows the normalized number, volume, and 
projected-surface-area distributions of our Martian analogue 
palagonite particles as functions of $\log r$, with $r$ the radius 
of a projected-surface-area equivalent sphere 
\citep[for details on these size
distributions, see Appendix A of][]{2005JQSRT..90..191V}. 
The number distribution of our palagonite particles was 
approximated by a log-normal distribution, yielding an effective radius, 
$r_{\rm eff}$, of 4.46~$\mu$m 
and an effective variance, $v_{\rm eff}$, of 7.29.
Note that $v_{\rm eff}$ is a dimensionless parameter. 
For precise definitions of $r_{\rm eff}$ and $v_{\rm eff}$
see \citet{1974SSRv...16..527H}, Eqs. (2.53) and (2.54), respectively.

We are well aware of the fact that the sizes of real Martian dust particles 
can be very different from those in our sample. Indeed, 
sizes of dust particles on Mars will probably vary from location to location, and
from time to time, especially when in local or global storms, dust particles 
are lifted up from the surface to be deposited somewhere else:
depending on the atmospheric turbulence, the particles in the Martian
atmosphere could have very different size distributions than those on 
the surface.

The effective radius of 4.46~$\mu$m of our sample particles
is a factor of 2 to 3 larger than the values put forward 
for the effective radius of Martian dust by \cite{2003JGRE.108i....1W}, 
who analyzed observations by the Thermal Emission Spectrometer (TES)
on-board the Mars Global Surveyor, and by \citet{1995JGR...100.5235P}, 
who analyzed observations performed by the Viking Lander.
In particular, \citet{1995JGR...100.5235P} derived
an effective radius of 1.85~$\pm~0.3~\mu$m.
From Pathfinder measurements, \citet{1999JGR...104.8987T} 
derived an effective radius of $1.6 \pm 0.15$ $\mu$m.
\citet{2004Sci...306.1753L} derived values from observations by the
Mars Exploration Rovers Spirit and Opportunity that are similar to
those of \citet{1995JGR...100.5235P} and \citet{1999JGR...104.8987T}.

Although our sample particles thus seem to be rather large, 
it should be noted that particle sizes as derived from 
observations will depend on the observing method, e.g. looking at
diffuse skylight or at the surface, as well as on the retrieval method.
In particular, according to numerical similations by \citet{2002SoSyR..36..367D}, effective radii
that are derived for spheroidal dust particles at visibile wavelengths,
under the assumption that these particles are spherical, can be
significantly underestimated. At infrared wavelengths,
\citet{2003A&A...404...35M} and \citet{2001A&A...378..228F} show that absorption and emission
processes, even in the small size parameter regime (i.e. $2 \pi r_{\rm eff}/\lambda \leq  1$),  
depend on the particle shape, too.
Clearly, because Martian dust is expected to show a great
variety in microphysical properties,
our results should simply be regarded as an example of what can be
expected for the scattering properties of irregularly shaped particles.

\section{The scattering matrix}
\label{section_scatteringmatrix}

\subsection{Definition of the scattering matrix}
\label{section_definitionmatrix}

The flux and state of  polarization of a quasi-monochromatic
beam of light can be described by means of a so-called flux vector. 
If such a beam of light is scattered by an ensemble of randomly oriented 
particles, separated by distances larger than 
their linear dimensions and in the absence of multiple scattering as 
in our experimental set-up
(see Sect.~\ref{section_setup}), 
the flux vectors of
the incident beam, $\pi{\bf\Phi_{0}}(\lambda)$, and scattered beam,
$\pi{\bf\Phi}(\lambda,\Theta)$, are for each scattering direction, 
related by a $4 \times 4$ matrix, as follows
\citep[][]{1957lssp.book.....V,2006JQSRT.100..437V}:
\begin{equation}
\label{matrix}
{\bf\Phi}(\lambda, \Theta)
	 = \frac{\lambda^{2}}{4\pi^{2}D^{2}}
	   \left(  \begin{array}{c c c c}
		  F_{11}&F_{12}& F_{13} & F_{14}\\
		  F_{12}&F_{22}& F_{23} & F_{24}\\
	         -F_{13}&-F_{23}& F_{33} & F_{34}\\
		  F_{14}&F_{24}& -F_{34} & F_{44} \\
		  \end{array} \right) {\bf\Phi_{0}}(\lambda),
\label{eq_scatmat}
\end{equation}
where the first elements of the column vectors are
fluxes divided by $\pi$ and the other elements describe the state 
of polarization of the beams by means of Stokes parameters.
Furthermore, $\lambda$ is the wavelength, and $D$ is the 
distance between the ensemble of particles and the detector. 
The scattering plane, i.e. the plane containing the directions 
of the incident and scattered beams, is the plane of reference 
for the flux vectors. 
The matrix, ${\bf F}$, with elements $F_{ij}$ is called the 
scattering matrix of the ensemble.
The scattering matrix elements $F_{ij}$ 
are dimensionless, and depend on the number of the particles and on their microphysical properties
(size, shape and refractive index), the wavelength 
of the light, and the scattering direction. 
For randomly oriented particles, the scattering
direction is fully described by the scattering angle $\Theta$,
the angle between the directions of propagation of the
incident and the scattered beams.

According to Eq.~(\ref{eq_scatmat}), a scattering matrix
has in general 10 different matrix elements.
For randomly oriented particles with equal 
amounts of particles and their mirror particles, as we
can assume applies for the particles of our ensemble, 
the four elements $F_{13}(\Theta)$, $F_{14}(\Theta)$,
$F_{23}(\Theta)$, and $F_{24}(\Theta)$ are zero over 
the entire scattering angle range \citep[see][]{1957lssp.book.....V}.
This leaves us only six non-zero scattering matrix
elements, as follows
\begin{equation}
   {\bf F}(\Theta)= \left[ \begin{array}{cccc}
      F_{11}(\Theta) & F_{12}(\Theta) & 0 & 0 \\
      F_{12}(\Theta) & F_{22}(\Theta) & 0 & 0 \\
      0 & 0 &  F_{33}(\Theta) & F_{34}(\Theta) \\
      0 & 0 & -F_{34}(\Theta) & F_{44}(\Theta)
      \end{array} \right],
\label{eq_scatteringmatrix}
\end{equation}
where $|F_{ij}(\Theta)/F_{11}(\Theta)| \leq 1$ 
\citep[][]{1986A&A...157..301H}.

For unpolarized incident light, matrix element $F_{11}(\Theta)$ is 
proportional to the flux of the singly scattered light and is also called the phase function. 
Also, for unpolarized incident light, 
the ratio $-F_{12}(\Theta)/F_{11}(\Theta)$ equals the 
degree of linear polarization of the scattered light.
The sign indicates the direction of polarization: 
a negative degree of polarization indicates that the 
scattered light is polarized parallel to the reference plane, whereas 
a positive degree of polarization indicates that the light is polarized perpendicular to the
reference plane. In calculations for fluxes only and where light 
is scattered only once, $F_{11}(\Theta)$ is the only matrix element 
that is required. Ignoring the other matrix elements, and hence the state 
of polarization of the light, in multiple 
scattering calculations, usually leads to errors in calculated fluxes 
\citep[see e.g.][]{1998GeoRL..25..135L,2002Icar..156..474M,2005A&A...444..275S}. 

\subsection{The experimental set-up}
\label{section_setup}

\begin{figure}
\vspace*{5.0cm}
\begin{center}
\includegraphics[width=17.0cm]{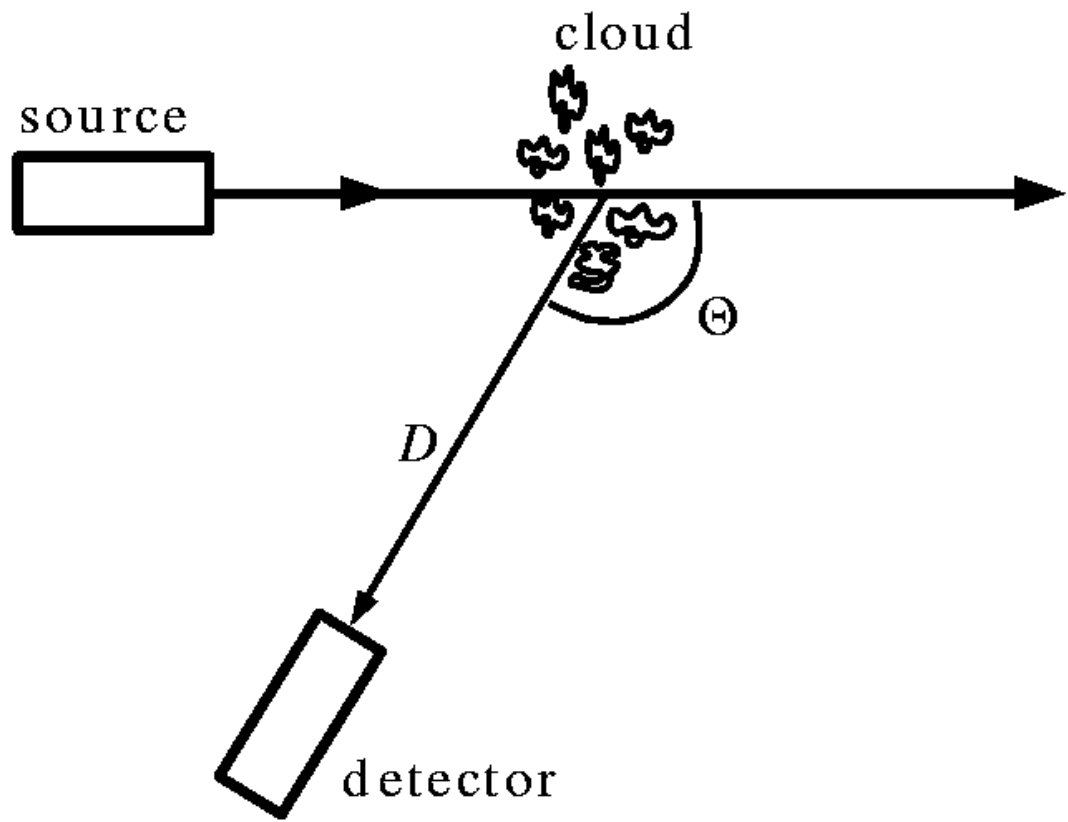}
\end{center}
\caption{Sketch of the experimental set--up. The light source (a HeNe laser) 
        emits a beam of quasi-monochromatic light, which passes through 
        a polarizer and a modulator (not shown).
	The light is scattered by the ensemble of palagonite particles. 
	A detector measures the flux that is scattered over
	a scattering angle $\Theta$. Measurements can be 
	performed for $3^\circ \leq \Theta \leq 174^\circ$.}  
\label{fig_setup}
\end{figure}

Our measurements have been performed with the light scattering 
experiment located in Amsterdam, the Netherlands
\citep[see e.g.][]{2005JQSRT..90..191V,
2003JQSRT..79..741H,Mishchenko2000_hovenier,
2001JGR...10617375V,2000A&A...360..777M}.  
Figure~\ref{fig_setup} shows a sketch of the experimental set--up.
In our experimental apparatus, we use a HeNe laser ($\lambda$=632.8 nm, 5 mW) 
as a light source. The laser light passes through a polarizer and an electro-optic 
modulator. The modulated light is subsequently scattered by an ensemble of 
randomly oriented particles from the sample, located in a 
jet stream produced by an aerosol generator.
The scattered light may pass through a quarter-wave plate and an 
analyzer, depending on the scattering matrix element of interest 
\citep[for details see e.g.][]{2005JQSRT..90..191V}, 
and is then detected by a photomultiplier 
tube which moves in steps along a ring with radius $D$ (see Eq.~(\ref{eq_scatmat})) 
around the ensemble of particles; in this way a range of scattering angles 
from 3$^\circ$ (nearly forward scattering) to 174$^\circ$
(nearly backward scattering) is covered in the measurements. 
 
We cannot measure close ($< 3^\circ$) to the exact forward scattering
direction, because there our detector would intercept the unscattered
part of the incident beam, nor can we measure close ($< 6^\circ$) to the 
exact backscattering direction, because there our detector would 
interfere with the incoming beam of light.

A photomultiplier placed at a fixed position (i.e. at a fixed scattering angle)
is used to correct the measured scattered fluxes for time fluctuations in 
the particle stream. It can safely 
be assumed that during the measurements, the particles are in the 
single scattering regime \citep[][]{2003JQSRT..79..741H}.

Due to the lack of measurements between 0$^\circ$ and 3$^\circ$
and between 174$^\circ$ and 180$^\circ$, we cannot measure the absolute 
angular dependency of the phase function, e.g. normalized to unity when
averaged over all scattering directions. Instead, we normalize the
measured phase function to unity at a scattering angle of 30$^\circ$.
We present the other scattering matrix elements divided by
the original measured phase function. We thus present ratios of 
elements of the scattering matrix instead of the elements themselves.  

\subsection{Measurements}
\label{section_measuredscatteringmatrix}

\begin{figure}
\begin{center}
\vspace*{-6.0cm}
\hspace*{-2.0cm}
\includegraphics[width=16.0cm]{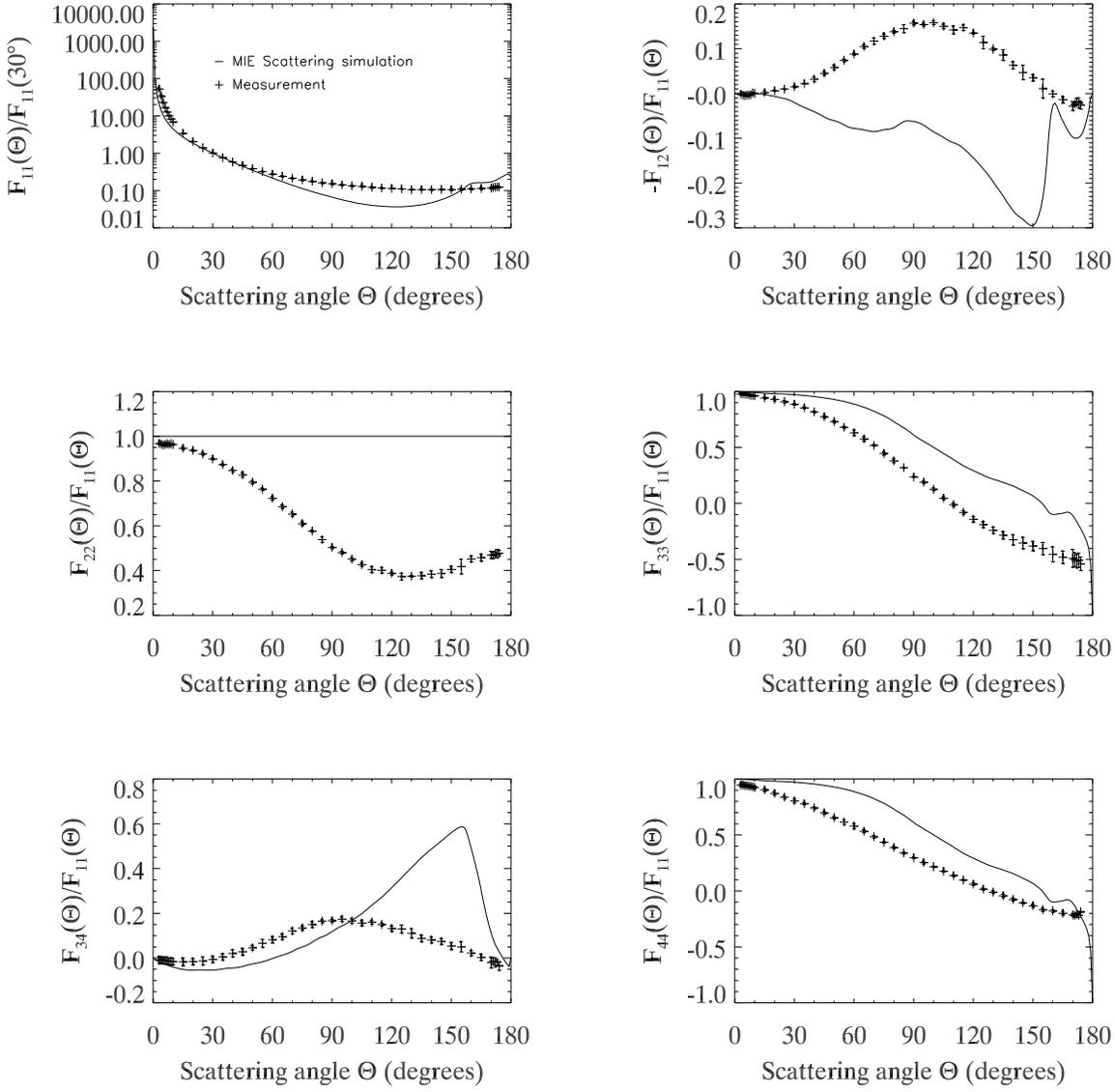}
\vspace*{-0.0cm}
\end{center}
\caption{Ratios of scattering matrix elements as functions of the scattering angle.
        Crosses show the ratios of elements measured, at $\lambda=632.8$~nm, 
	for the Martian analogue palagonite particles,
        with the vertical bars indicating the experimental errors.
        Solid lines show the ratios of elements calculated using Mie-theory for
        homogeneous, spherical particles that are
        distributed in size as shown in Fig.~\ref{fig_size}.
        }
\label{fig_matrix}
\end{figure}

Figure~\ref{fig_matrix} shows the six measured not identically zero 
(cf. Eq.~\ref{eq_scatteringmatrix}) ratios of elements of the 
scattering matrix of the Martian analogue palagonite particles as functions 
of the scattering angle $\Theta$, together with the experimental errors.
We have verified that the measured ratios of the elements of the scattering 
matrix satisfy the Cloude coherency matrix test 
\citep[][]{2004Hovenier} within the experimental errors.
And we verified that the other measured ratios of the elements of the
scattering matrix, i.e. $F_{13}(\Theta)/F_{11}(\Theta)$, $F_{23}(\Theta)/F_{11}(\Theta)$,
$F_{14}(\Theta)/F_{11}(\Theta)$, and $F_{24}(\Theta)/F_{11}(\Theta)$,
do not differ from zero by more than the experimental errors
(see Eq.~(\ref{eq_scatteringmatrix})).

To illustrate the influence of the particle shape on the scattering
behavior of the palagonite particles, the measurements in Fig.~\ref{fig_matrix} 
are presented together with results of Lorenz-Mie calculations 
\citep[][]{1957lssp.book.....V,1984A&A...131..237D} for homogeneous, 
optically nonactive, spherical particles at a wavelength of 632.8~nm. 
For the Lorenz-Mie calculations we employed the number size distribution, 
$n(r)$, derived from the measured projected-surface-area distribution,
and the refractive index was fixed to  $m=1.5+0.0005i$ 
(cf. Sect.~\ref{section_palagonite} and Fig.~\ref{fig_size}).

As can be seen in Fig.~\ref{fig_matrix}, the measured phase function, i.e.
$F_{11}(\Theta)/F_{11}(30^\circ)$, of the irregularly shaped Martian 
analogue palagonite particles covers almost three orders of 
magnitude between $\Theta=3^\circ$ and $\Theta=174^\circ$, 
with a strong peak towards the smallest scattering angles (the
so-called forward scattering peak) and a smooth drop-off towards 
the largest scattering angles. The measured phase 
function is very flat for scattering over intermediate
($70^\circ < \Theta < 150^\circ$) and large ($\Theta > 150^\circ$) scattering angles.
The relatively flat appearance of the phase function of the palagonite
particles at large scattering angles appears to be a general behavior for 
(terrestrial) irregularly shaped mineral particles with moderate 
refractive indices 
\citep[see e.g.][]{2001JGR...10617375V,2000A&A...360..777M,
2001JGR...10622833M}.
Our palagonite phase function resembles the phase functions measured in-situ 
with the Viking \citep[][]{1995JGR...100.5235P} and Pathfinder missions
\citep[][]{1999JGR...104.8987T}. 
In Sect.~\ref{section_summary}, we make a more detailed comparison between
our phase function and those presented by \citet{1999JGR...104.8987T}.

As mentioned before (see Sect.~\ref{section_definitionmatrix}),
the ratio $-F_{12}(\Theta)/F_{11}(\Theta)$ represents the degree 
of linear polarization of the singly scattered light for incident 
unpolarized light. For the irregularly shaped palagonite particles, 
Fig.~\ref{fig_matrix} shows that this ratio has a characteristic 
(positive) bell shape at intermediate scattering
angles and a small negative branch for $\Theta \gtrsim 160^\circ$.
For scattering angles larger than about 140$^\circ$, the 
scattering angle dependence of our measured ratio
$-F_{12}(\Theta)/F_{11}(\Theta)$ resembles Earth-based 
observations of the planetary phase angle dependence of the
degree of linear polarization of Mars
\citep[][]{2003SoSyR..37...87D,2005Icar..176....1S}
(the planetary phase angle equals 180$^\circ - \Theta$
for single scattering). 
This suggests that the polarization opposition effect 
that is observed at small phase angles for most solid solar 
system bodies
\citep[see e.g.][and references therein]{2005Icar..179..490R}
can be explained, at least partly, by single scattering by small 
irregular particles. 
Here, it should be noted that the observations discussed by 
\citet{2003SoSyR..37...87D} and \citet{2005Icar..176....1S} 
pertain to light that has been scattered in the Martian
atmosphere combined with light that has been reflected by the surface.
It is thus not purely representative for airborne dust particles.

The most striking difference between the measured and the calculated ratios  
$-F_{12}(\Theta)/F_{11}(\Theta)$ (Fig.~\ref{fig_matrix}) 
is their sign, hence the direction of polarization of the scattered 
light for unpolarized incident
light. The irregularly shaped particles mostly yield scattered light
polarized perpendicular to the reference plane, while the 
spherical particles yield scattered light polarized parallel to this plane.
Another difference is that for the irregularly shaped particles,
ratio $-F_{12}(\Theta)/F_{11}(\Theta)$ is a smooth, almost
featureless function of $\Theta$, while for the spherical
particles, the ratio shows strong angular features, 
especially at large scattering angles.
 
Scattering matrix element ratio $F_{22}(\Theta)/F_{11}(\Theta)$ 
is often used as a measure for the non-sphericity of the scattering
particles, since for homogenous, optically inactive spheres, 
this ratio equals unity at all scattering angles.
As can be seen in Fig.~\ref{fig_matrix}, for the irregularly 
shaped palagonite particles, 
$F_{22}(\Theta)/F_{11}(\Theta)$ deviates significantly from unity at
all but the smallest scattering angles. Indeed, with increasing scattering
angle, it decreases to slightly
below 0.4 at $\Theta \approx 130^\circ$, and then increases
again to 0.5 when $\Theta$ approaches 180$^\circ$. 
The scattering angle dependence measured for the palagonite
particles is similar in shape to that reported for irregularly shaped 
mineral aerosol particles \citep[][]{2001JGR...10617375V},
and for e.g. various types of volcanic ashes
\citep[][]{2004JGRD..10916201M}. According to 
\citet{2001JGR...10617375V}, the minimum value at intermediate
scattering angles and the maximum value at the largest scattering
angles are affected by the size and refractive index of the
particles.

Another indication for the shape of the scattering particles 
are the ratios $F_{33}(\Theta)/F_{11}(\Theta)$ and
$F_{44}(\Theta)/F_{11}(\Theta)$. As can also be seen in 
Fig.~\ref{fig_matrix}, for homogenous, optically 
inactive spheres, $F_{33}(\Theta) \equiv F_{44}(\Theta)$
\citep[][]{2004Hovenier},
whereas we find significant differences between the 
measured $F_{44}(\Theta)/F_{11}(\Theta)$ and
$F_{33}(\Theta)/F_{11}(\Theta)$ for the palagonite sample. 
The ratio $F_{33}(\Theta)/F_{11}(\Theta)$ is zero at a smaller
scattering angle than $F_{44}(\Theta)/F_{11}(\Theta)$,
and has a lower minimum (-0.5 versus -0.2).
Indeed, for the irregularly shaped particles, 
these ratios show an apparently typical behavior
for non-spherical particles \citep[][]{Mishchenko2000}, namely,
at large scattering angles, $F_{44}(\Theta)/F_{11}(\Theta)$
is larger than $F_{33}(\Theta)/F_{11}(\Theta)$.

Finally, scattering matrix element ratio 
$F_{34}(\Theta)/F_{11}(\Theta)$ of the irregularly shaped
particles shows a shallow bell shape 
with slightly negative branches for $\Theta < 30^\circ$ 
and for $\Theta > 165^\circ$.
This scattering angle dependence is commonly found 
for irregularly shaped silicate particles 
\citep[e.g.][]{2005JQSRT..90..191V,2001JGR...10617375V,
2000A&A...360..777M}.
Interestingly, whereas for the irregularly shaped particles, 
$F_{34}(\Theta)/F_{11}(\Theta)$ is very similar to
$-F_{12}(\Theta)/F_{11}(\Theta)$, for the spherical particles,
these ratios differ strongly from each other, both in sign and
in shape, as can be seen in Fig.~\ref{fig_matrix}.

Comparison between the measured and the calculated scattering matrix 
element ratios in Fig.~\ref{fig_matrix} supports the idea that 
scattering by non-spherical particles generally leads to
smoother functions of the scattering angle than
scattering by spherical particles. This smooth scattering behavior
by irregularly shaped particles proves to be very difficult 
to simulate numerically without taking into account 
the irregular shape of the particles 
\citep[][]{2003JGRD.108aAAC12N,2003JQSRT..79.1031N,
2003JPhD...36..915K}.

An electronic table of the measured ratios of the elements of the 
scattering matrix will be available from the Amsterdam Light Scattering 
Database \citep[][]{2005JQSRT..90..191V,2006JQSRT.100..437V}.

\subsection{The auxiliary scattering matrix}
\label{section_auxiliarymatrix}

It appears to be difficult to directly use the measured ratios of elements of the scattering
matrix in radiative transfer calculations, because of 
the lack of measurements below $\Theta=3^\circ$ and above 
$\Theta=174^\circ$. 
In particular, it would be interesting to have the forward scattering
peak in the phase function since it contains a large fraction of the 
scattered energy (see Fig.~\ref{fig_matrix}), and is thus very
important for the accurate modelling of scattered light
in e.g. planetary atmospheres.
In addition, the lack of measurements at small and large scattering
angles inhibits the normalization of scattering matrix elements 
such that the average of the phase function over all scattering
directions equals unity.
With such a normalization and a value
for the single scattering albedo of the scattering particles,
one could model the absolute amount of radiation that is
scattered in a given direction.

To facilitate the use of the measured ratios of elements of the scattering 
matrix in radiative transfer calculations, we construct from these a so--called {\em auxiliary
scattering matrix}, ${\bf F^{\rm au}}$, for which holds (for $i,j=1$ to 4 with the exception of $i=j=1$), 
\begin{equation}
   F^{\rm au}_{ij}(\Theta) = \frac{F_{ij}(\Theta)}{F_{11}(\Theta)} F_{11}^{\rm au}(\Theta),
\label{equation_auxiliary}
\end{equation}
where the auxiliary phase function ${F_{11}^{\rm au}}$ is equal to
\begin{equation}
    F_{11}^{\rm au}(\Theta) =
    \frac{F_{11}(\Theta)}{F_{11}(30^{\circ})} F_{11}^{\rm au}(30^{\circ}) 
    \hspace*{1cm} {\rm for} \hspace*{0.5cm}
    3^\circ \leq \Theta \leq 174^\circ.
\label{ratios}
\end{equation}

This auxiliary phase function is normalized 
according to
\begin{equation}
   \frac{1}{4\pi} \int_{4\pi} F^{\rm au}_{11}(\Theta) \, d\omega = 1,
\label{eq_normalization}
\end{equation}
where $d \omega$ is an element of solid angle. 
Combining Eq.~(\ref{ratios}) and Eq.~(\ref{eq_normalization}) and 
setting $F^{\rm au}_{11}(30^\circ)$ equal to $1/C$ leads to 
\begin{equation}
   \frac{1}{4\pi} \int_{4\pi} 
   \frac{F_{11}(\Theta)}{F_{11}(30^\circ)} \, d\omega = C.
\label{eq_normalization2}
\end{equation}
where $C$ is a normalization constant.
This constant can in principle be obtained by 
evaluating the integral on the left-hand side of Eq. ~(\ref{eq_normalization2}), 
provided the function to be integrated is known over the full range of scattering angles. 
Therefore, we added artificial datapoints at 0 and 180 degrees to the measured values 
of $F_{11}(\Theta)/F_{11}(30^\circ)$. At $\Theta=180^\circ$, 
the smoothness of the measured phase function allows us to simply add 
an artificial data point to $F_{11}(\Theta)/F_{11}(30^\circ)$ 
by spline extrapolation \citep[][]{1992nrfa.book.....P} 
of the measured data points. 

Adding an artificial data point to the measured $F_{11}(\Theta)/F_{11}(30^\circ)$ 
at $\Theta=0^\circ$, is more complicated. Numerical tests with the calculated phase 
function for the hypothetical homogeneous, spherical palagonite particles 
(see Fig.~\ref{fig_matrix}) show that extrapolation of the 
calculated phase function at $\Theta \leq 3^\circ$ towards 
$\Theta=0^\circ$, using e.g. splines \citep[][]{1992nrfa.book.....P}, 
fails to reproduce the strength of the calculated forward scattering 
peak. We thus decided not to extrapolate the measured phase function 
from $\Theta=3^\circ$ towards $\Theta=0^\circ$. Instead we add an artificial 
data point to the measured $F_{11}(\Theta)/F_{11}(30^\circ)$ at $\Theta=0^\circ$ using 
the phase function that 
we calculated for the projected-surface-area equivalent, homogeneous, 
spherical particles.
The rationale for this approach, which is similar to that used by 
\citet{2003JQSRT..79..903L} and \citet{2007JGRD..11213215M}, 
is that the forward scattering peak results mainly from the 
diffraction of the incident 
light. The strength of the diffraction peak and its scattering
angle dependence appears to depend mainly on the size of 
the particles and fairly shape independent for 
projected-surface-area equivalent convex particles in random orientation 
\citep[][]{2002sael.book.....M}. 

Because our normalization of the measured phase function, $F_{11}(\Theta)/F_{11}(30^\circ)$, at 
$\Theta=30^\circ$ is rather arbitrary (we could have chosen
a different value of $\Theta$ for the normalization), we scale the phase function
as calculated for the spherical palagonite particles to the measured phase function. 
For this, we use the following equation,
\begin{equation}
      \frac{F_{11}(0^\circ)}{F_{11}(30^\circ)} = 
      \frac{F_{11}^{\rm s}(0^\circ)}{F_{11}^{\rm s}(3^\circ)}
      \frac{F_{11}(3^\circ)}{F_{11}(30^\circ)},
\label{eq_exp200}
\end{equation}
with $F_{11}(0^\circ)/F_{11}(30^\circ)$ the artificial data point at 
$\Theta=0^\circ$ and the superscript "s" indicating the phase function as calculated 
for the spherical particles.

We now have data points available across the full scattering angle 
range to evaluate the integral in Eq.~(\ref{eq_normalization2})
and to obtain the normalization constant $C$.
The numerical evaluation of this integral, however, appears to be
difficult because of the steep slope between $\Theta=0^\circ$ and 
$\Theta=3^\circ$, where no measured data points are available.
Therefore, we have chosen a different method to obtain the normalization 
constant $C$ in Eq.~(\ref{eq_normalization2}).
This method is based on the expansion of the measured phase function 
$F_{11}(\Theta)/F_{11}(30^\circ)$, including the added, artificial, 
data points at $\Theta=0^\circ$ to $\Theta=180^\circ$, 
as a function of the scattering angle into so--called generalized spherical 
functions \citep[][]{1963Gelfand,1983A&A...128....1H,1984A&A...131..237D,
2004Hovenier}. This method of obtaining expansion coefficients from scattering 
matrix elements is explained in detail in Sect.~\ref{section_expansioncoefs}.
The expansion of $F_{11}(\Theta)/F_{11}(30^\circ)$ yields expansion coefficients 
$\alpha^l_1$ (with $0 \leq l$). The first of these expansion coefficients, 
$\alpha^0_1$, is equal to constant $C$ in Eq.~(\ref{eq_normalization2}) \citep[][]{2004Hovenier}. 
Having obtained $C$ in this way, and thus $F_{11}^{\rm au}(30^{\circ})$, 
we readily find $F_{11}^{\rm au}(\Theta)$ from 
the measured ratio $F_{11}(\Theta)/F_{11}(30^\circ)$ and Eq. ~(\ref{ratios}). 

Next, given the auxiliary phase function, we derive the synthetic matrix elements 
$F_{ij}^{\rm au}(\Theta)$ for $i,j=1$ to 4 with the exception of $i=j=1$ from 
the measured ratios $F_{ij}(\Theta)/F_{11}(\Theta)$, using Eq.~(\ref{equation_auxiliary}). 
To obtain also the complete scattering angle range
for the other scattering matrix element ratios, we extrapolate the 
measured $F_{ij}(\Theta)/F_{11}(\Theta)$ ($i,j=1$ to 4 with the exception of $i=j=1$) 
towards $\Theta=0^\circ$ and $180^\circ$.
At these two scattering angles, the following equalities should hold:
\citep[see Display 2.1 in][]{2004Hovenier}
\begin{eqnarray}
  & & F_{12}(0^\circ)/F_{11}(0^\circ) = F_{34}(0^\circ)/F_{11}(0^\circ) = 0, \label{eq6} \\
  & & F_{22}(0^\circ)/F_{11}(0^\circ) = F_{33}(0^\circ)/F_{11}(0^\circ), \label{eq5} \\
  & & F_{22}(180^\circ)/F_{11}(180^\circ) = -F_{33}(180^\circ)/F_{11}(180^\circ), \label{eq7} \\
  & & F_{12}(180^\circ)/F_{11}(180^\circ) = F_{34}(180^\circ)/F_{11}(180^\circ) = 0, \label{eq8} \\
  & & F_{44}(180^\circ)/F_{11}(180^\circ) = 1 - 2 F_{22}(180^\circ)/F_{11}(180^\circ). \label{eq9}
\end{eqnarray}
Following Eq.~(\ref{eq6}), ratios $F_{12}(0^\circ)/F_{11}(0^\circ)$ and 
$F_{34}(0^\circ)/F_{11}(0^\circ)$ are set equal to zero. Following Eq.~(\ref{eq5}), 
we use splines to extrapolate the ratios $F_{22}(\Theta)/F_{11}(\Theta)$
and $F_{33}(\Theta)/F_{11}(\Theta)$ towards $\Theta=0^\circ$,
and we set both $F_{22}(0^\circ)/F_{11}(0^\circ)$ and $F_{33}(0^\circ)/F_{11}(0^\circ)$
equal to the average of the two extrapolated values. 
Ratio $F_{44}(0^\circ)/F_{11}(0^\circ)$ is obtained
by extrapolating (with splines) the ratio $F_{44}(\Theta)/F_{11}(\Theta)$
from $\Theta=3^\circ$ towards $\Theta=0^\circ$.

In the backward scattering direction, the measured scattering matrix
element ratios $F_{ij}(\Theta)/F_{11}(\Theta)$ appear to be smooth 
functions of $\Theta$ (see Fig.~\ref{fig_matrix}). We use splines
\citep[][]{1992nrfa.book.....P} to extrapolate 
$F_{22}(\Theta)/F_{11}(\Theta)$, and $F_{33}(\Theta)/F_{11}(\Theta)$ 
from $\Theta=174^\circ$ to $\Theta=180^\circ$.
Because $F_{22}(180^\circ)/F_{11}(180^\circ)$ should be equal to 
$-F_{33}(180^\circ)/F_{11}(180^\circ)$ (see Eq.~(\ref{eq7})), we set both 
$F_{22}(180^\circ)/F_{11}(180^\circ)$ and 
$-F_{33}(180^\circ)/F_{11}(180^\circ)$ equal to the average of the two 
extrapolated values.
Following Eq.~(\ref{eq8}), we set $F_{12}(180^\circ)/F_{11}(180^\circ)$ and 
$F_{34}(180^\circ)/F_{11}(180^\circ)$ equal to zero,
and calculate $F_{44}(180^\circ)/F_{11}(180^\circ)$ using Eq.~(\ref{eq9})
with $F_{22}(180^\circ)/F_{11}(180^\circ)$.

In the following, we will refer to the elements of our auxiliary
scattering matrix ${\bf F}^{\rm au}(\Theta)$ as
\citep[see][]{2004Hovenier}:
\begin{equation}
   {\bf F}^{\rm au}(\Theta)= \left[ \begin{array}{cccc}
      a_1(\Theta) & b_1(\Theta) & 0 & 0 \\
      b_1(\Theta) & a_2(\Theta) & 0 & 0 \\
      0 & 0 &  a_3(\Theta) & b_2(\Theta) \\
      0 & 0 & -b_2(\Theta) & a_4(\Theta)
      \end{array} \right].
\label{eq_scatteringmatrixS}
\end{equation}

\section{Expansion coefficients}
\label{section_expansioncoefs}

\subsection{Definitions of the expansion coefficients}

For use in numerical radiative transfer algorithms, it is often
advantageous to expand the elements of a scattering matrix
as functions of the scattering angle into so--called
generalized spherical functions
\citep[][]{1963Gelfand,1983A&A...128....1H,1984A&A...131..237D,
2004Hovenier}.
The advantage of using the coefficients of this expansion, 
the so--called expansion coefficients, instead 
of the elements of a scattering matrix themselves is that it 
can significantly speed up multiple scattering calculations,
which is of particular importance when polarization is taken 
into account \citep[][]{2004Hovenier,1987A&A...183..371D}.

We indicate the generalized spherical functions by
$P_{m,n}^l(\cos{\Theta})$ with the indices $m$ and $n$  
equal to +2, +0, -0, or -2,
and with $l \geq {\rm max} \{ |m|,|n| \}$. 
Note that generalized spherical function $P_{0,0}^l$ is 
simply a Legendre polynomial.
The expansion of the elements of auxiliary scattering
matrix ${\bf F}^{\rm au}(\Theta)$ (see Eq.~(\ref{eq_scatteringmatrixS}))
into generalized spherical functions is as follows: 
\begin{eqnarray}
   a_1(\Theta) &=&
     \sum_{l=0}^\infty \alpha_1^l P_{0,0}^l(\cos{\Theta}),
     \label{eq_exp1} \\
   a_2(\Theta) + a_3(\Theta) &=&
     \sum_{l=2}^\infty (\alpha_2^l + \alpha_3^l) P_{2,2}^l(\cos{\Theta}), 
     \label{eq_exp2} \\
   a_2(\Theta) - a_3(\Theta) &=&
     \sum_{l=2}^\infty (\alpha_2^l - \alpha_3^l) P_{2,-2}^l(\cos{\Theta}),
     \label{eq_exp3} \\
   a_4(\Theta) &=&
     \sum_{l=0}^\infty \alpha_4^l P_{0,0}^l(\cos{\Theta}),
     \label{eq_exp4} \\
   b_1(\Theta) &=&
     \sum_{l=2}^\infty \beta_1^l P_{0,2}^l(\cos{\Theta}),
     \label{eq_exp5} \\
   b_2(\Theta) &=&
     \sum_{l=2}^\infty \beta_2^l P_{0,2}^l(\cos{\Theta}).
     \label{eq_exp6}
\end{eqnarray}
Here, $\alpha_1^l$, $\alpha_2^l$, $\alpha_3^l$, $\alpha_4^l$, 
$\beta_1^l$, and $\beta_2^l$ are the expansion coefficients.
For each value of integer $l$, the expansion coefficients can be 
derived from the auxiliary scattering matrix elements 
using the definitions of the generalized spherical functions 
and their orthogonality relations \citep[see][]{2004Hovenier}.

A similar expansion in generalized spherical functions can be made 
for any scattering matrix of the form given by Eq. ~(\ref{eq_scatteringmatrix}). 
The coefficient $\alpha^0_1$ is always equal to the average of the one-one element 
over all directions \citep[][]{2004Hovenier}. 
So for the auxiliary phase function $a_1(\Theta)$ we have, 
according to Eq. ~(\ref{eq_normalization}), $\alpha^0_1$ = 1 and for the measured phase 
function, $F_{11}(\Theta)/F_{11}(30^{\circ})$ we have $\alpha^0_1$ = C, 
as mentioned in Sect.~\ref{section_auxiliarymatrix}.
\subsection{The Singular Value Decomposition method}

To derive the expansion coefficients of the measured phase function, 
incuding the points added at 0 and 180 degrees, 
and of all six elements of the auxiliary scattering matrix we write each of
the Eqs.~(\ref{eq_exp1})--(\ref{eq_exp6}) 
into the following general form:
\begin{equation}
   y(\Theta) = \sum_{l=n}^{m} \gamma^l X^l(\Theta),
\label{eq_exp7}
\end{equation}
where $y(\Theta)$ represents the value of 
the measured phase function or an auxiliary scattering matrix 
element at scattering angle $\Theta$, or,
in the case of $a_2$ and $a_3$, respectively 
their sum, as in Eq.~(\ref{eq_exp2}), or difference, as in
Eq.~(\ref{eq_exp3}). 
The functions $X^l(\Theta)$ in Eq.~(\ref{eq_exp7}) 
are the basis functions, 
for which we choose the appropriate generalized spherical functions
\citep[][]{2004Hovenier,1987A&A...183..371D}.
The parameters $\gamma^l$ in Eq.~(\ref{eq_exp7}) represent the 
expansion coefficients, or, in the case of 
$\alpha_2^l$ and $\alpha_3^l$, respectively
their sum (Eq.~(\ref{eq_exp2})) or difference (Eq.~(\ref{eq_exp3})).
Furthermore in Eq.~(\ref{eq_exp7}), $n$~equals~0 or~2, depending on
the auxiliary scattering matrix element under consideration
(see Eqs.~(\ref{eq_exp1})--(\ref{eq_exp6})), and, although
theoretically $m$~equals~$\infty$ 
(see Eqs.~(\ref{eq_exp1})--(\ref{eq_exp6})), in practice
$m$ is restricted to the number of scattering angles
at which values of the auxiliary scattering matrix elements 
$y(\Theta)$ are available. 

For a linear model such as that represented by Eq.~(\ref{eq_exp7}),
the merit function $\chi^2$ is generally defined as:
\begin{equation}
   \chi^2 = \sum_{i=1}^{k}\left[ 
            \frac{y(\Theta_i)-\sum_{l=n}^{m} \gamma^l 
            X^l(\Theta_i)}{\sigma_i} \right]^2,
\label{eq_exp100}
\end{equation}
where $k$ is the number of available data points and $\sigma_i$ 
is the error associated with data point $y(\Theta_i)$.
We use the Singular Value Decomposition (SVD) method 
\citep[][]{1992nrfa.book.....P} to solve Eq.~(\ref{eq_exp100})
for the expansion coefficients $\gamma^l$, because
this method is only slightly susceptible to roundoff errors and 
provides a solution that is the best approximation in
the least-squares sense, both for overdetermined systems (in which the
number of data points is larger than the required number of
expansion coefficients) and underdetermined systems (in which the
number of data points is smaller than the required number of 
expansion coefficients). 

To test the robustness and the quality of the fit method based
on the SVD method, we applied it to scattering matrix elements 
that we calculated for the hypothetical, spherical palagonite 
particles, and that are shown in Fig.~\ref{fig_matrix}
(note that our Mie-algorithm \citep[][]{1984A&A...131..237D} provides
matrix elements normalized according to 
Eq.~(\ref{eq_normalization}), although in Fig.~\ref{fig_matrix},
the normalization of the elements has been adapted to correspond to that
of the measurements). For the test
we compare the matrix elements calculated with our Mie-algorithm 
with the matrix elements obtained using the expansion
coefficients derived with the SVD method and 
Eqs.~(\ref{eq_exp1})-(\ref{eq_exp6}).  
For the relative errors in the matrix elements
we adopt the values of the experimental errors. 

We tested two aspects of the application of the SVD method. 
First, we applied the method to matrix elements calculated
at the same set of scattering angles as the measured ratios of matrix
elements, i.e. having a typical angular resolution of 5$^\circ$
and scattering angles ranging from $3^\circ$ to
174$^\circ$. Comparing the matrix elements obtained using
the expansion coefficients that were derived
with the SVD method with the directly calculated 
matrix elements, we found that the relatively coarse
angular sampling and the lack of data points below 
$\Theta=3^\circ$ gave rise to strong oscillations in the 
matrix elements that were calculated from the derived
expansion coefficients. In addition, with the derived expansion
coefficients, we could not reproduce the strong forward 
scattering peak in the phase function. 
The lack of data points above $\Theta=174^\circ$ appeared to 
be less of a problem, probably because of the smoothness 
of the matrix elements at those scattering angles.

Second, we applied the SVD method to matrix elements calculated
at an angular resolution of 1$^\circ$ and covering the full
scattering angle range, i.e. from 0$^\circ$ to 180$^\circ$. 
The matrix elements obtained using the hence derived expansion
coefficients coincided within the numerical precision with
the directly calculated matrix elements.
From this we conclude that our implementation of the SVD method
is reliable, but that we have to apply it to the whole
scattering angle range, and with a relatively high 
angular resolution. 

Finally, we found that averaging two sets of derived
expansion coefficients, one of which has one coefficient 
more than the other, 
removes most of the left-over oscillations in the scattering
matrix element that is calculated with the coefficients.

\subsection{The derived expansion coefficients}
\label{section_derivedcoefs}

For deriving the expansion coefficients of the scattering
matrix elements of the Martian analogue palagonite particles,
we use the elements of the auxiliary scattering matrix 
${\bf F}^{\rm au}(\Theta)$ (see Sect.~\ref{section_auxiliarymatrix}),
because they cover the whole scattering angle range and are normalized
according to Eq.~\ref{eq_normalization}.

We increase the angular sampling of the auxiliary scattering matrix by adding
artificial data points by spline interpolation to the dataset between $\Theta=3^\circ$ and $174^\circ$. 
Starting with 44 measured data points per matrix element, adding the
artificial data points results in 220 data points for each of the auxiliary scattering matrix elements.
This amount of artificial data points includes the values at the forward and backward scattering 
angles at respectively $\Theta=0^\circ$ and $\Theta=180^\circ$ as described in 
Sect.~\ref{section_auxiliarymatrix}.
The 220 datapoints proved to be required to be able to follow the steep slopes 
of $a_1(\Theta)$, $a_2(\Theta)$, $a_3(\Theta)$ and $a_4(\Theta)$ between
$\Theta=0^\circ$ and $\Theta=3^\circ$,
where no artifical datapoints were added, without introducing unwanted oscillations.  
As 220 datapoints are needed to obtain the auxilliary phase function $a_1(\Theta)$, 
it proved to be practical to extend also $F_{12}(\Theta)/F_{11}(\Theta)$ and 
$F_{34}(\Theta)/F_{11}(\Theta)$ to the same amount of 220 datapoints, 
to be able to straightforwardly apply Eq.~\ref{equation_auxiliary} to obtain 
$b_1(\Theta)$ and $b_2(\Theta)$.

\begin{figure}
\begin{center}
\vspace*{-6.0cm}
\hspace*{-2.0cm}
\includegraphics[width=16.0cm]{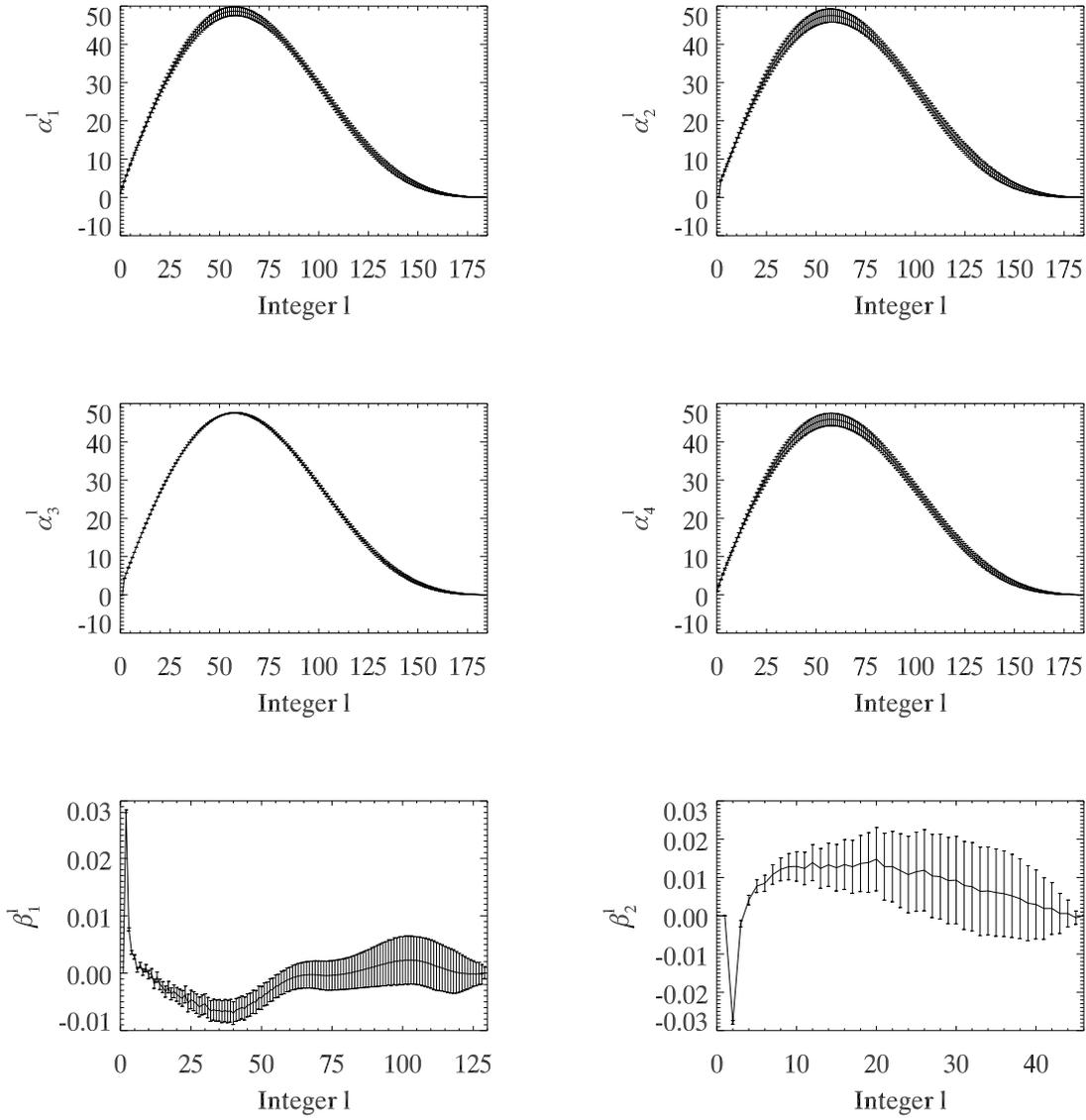}
\vspace*{-0.0cm}
\end{center}
\caption{The expansion coefficients $\alpha_1^l$, $\alpha_2^l$, 
         $\alpha_3^l$, $\alpha_4^l$, $\beta_1^l$ and $\beta_2^l$ as 
	 functions of integer $l$, with their absolute errors, as derived
         from the auxiliary scattering matrix elements $a_1(\Theta)$, 
         $a_2(\Theta)$, $a_3(\Theta)$, $a_4(\Theta)$, $b_1(\Theta)$,
         and $b_2(\Theta)$, respectively.}
\label{fig_expansioncoef}
\end{figure}

The optimal number of expansion coefficients for each of the elements
was obtained in an iterative process: unrealistic oscillations 
at large scattering angles are suppressed when using a smaller number of 
expansion coefficients, 
while the fit of the steep phase function near $0^\circ$ is improved when
using a larger number of expansion coefficients.
Applying our SVD method to the auxiliary scattering matrix elements, 
leaves us with respectively 185 expansion coefficients for $a_1(\Theta)$, 
$a_2(\Theta)$, $a_3(\Theta)$, $a_4(\Theta)$, 130 expansion 
coefficients for $b_1(\Theta)$ and 46 expansion coefficients for $b_2(\Theta)$.
Figure~\ref{fig_expansioncoef} shows 
the expansion coefficients $\alpha^l_1$, $\alpha^l_2$,
$\alpha^l_3$, $\alpha^l_4$, $\beta^l_1$, and $\beta^l_2$, 
derived from the auxiliary scattering matrix 
of the Martian analogue palagonite particles.
The expansion coefficients are plotted with error
bars that originate from the experimental errors in the measurements.
An electronic table of the expansion coefficients
will be available from the Amsterdam 
Light Scattering Database\footnote{Website: http://www.astro.uva.nl/scatter}
\citep[][]{2005JQSRT..90..191V,2006JQSRT.100..437V}.

\subsection{The synthetic scattering matrix}
\label{section_syntheticscatteringmatrix}

\begin{figure}
\begin{center}
\vspace*{-6.0cm}
\hspace*{-2.0cm}
\includegraphics[width=16.0cm]{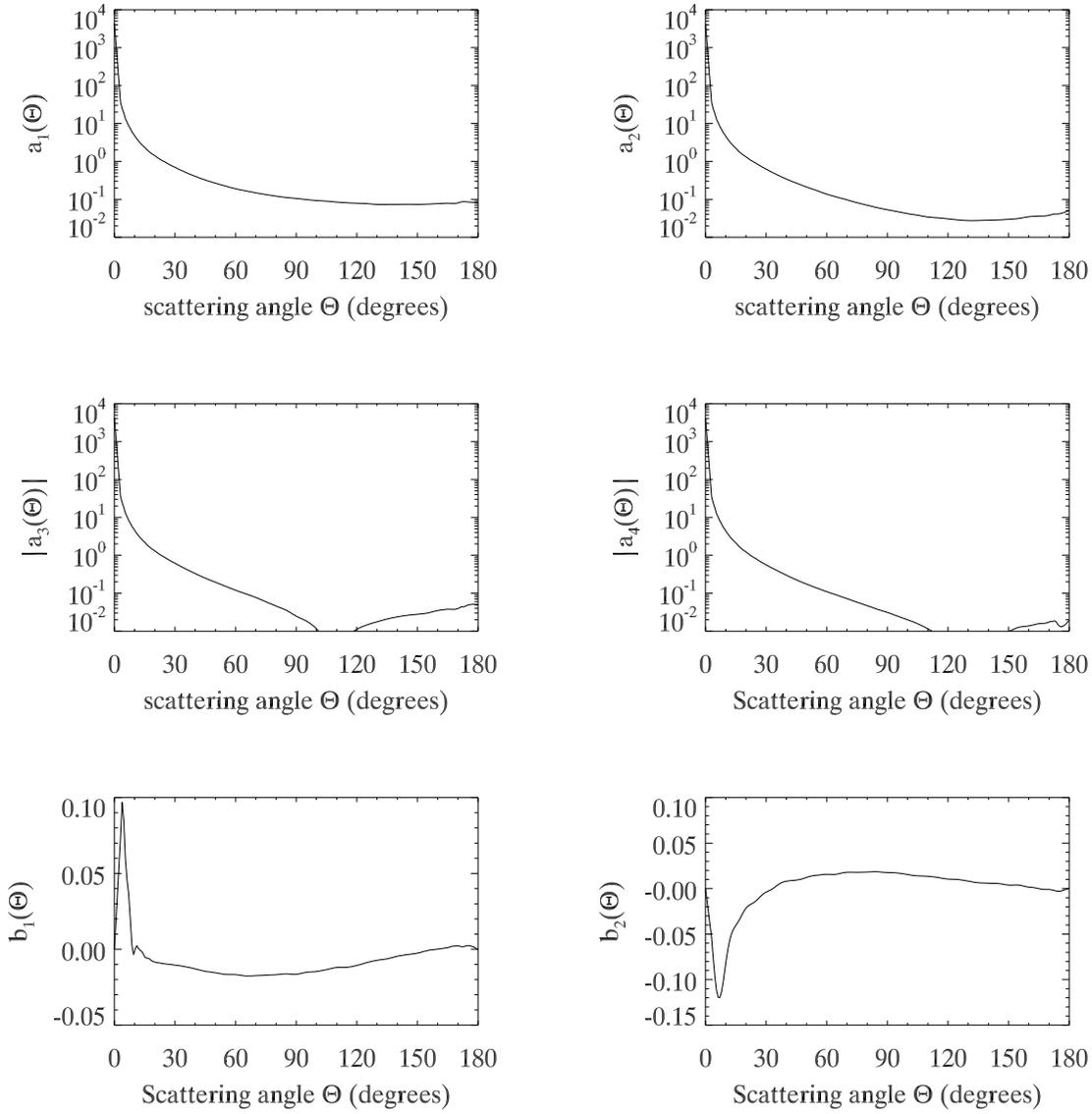}
\vspace*{-0.0cm}
\end{center}
\caption{The synthetic scattering matrix elements $a_1$, $a_2$, $a_3$, $a_4$,
         $b_1$, and $b_2$ as functions of the scattering angle $\Theta$.}
\label{fig_fitmeasurement}
\end{figure}

Employing Eqs.~(\ref{eq_exp1})--(\ref{eq_exp6}) with 
the expansion coefficients presented in Sect.~\ref{section_derivedcoefs}, 
we can now calculate, at an arbitrary angular resolution,
the so--called {\em synthetic scattering matrix} 
\citep[see also][]{2004JGRD..10916201M}
which covers the complete scattering range,
i.e. from 0$^\circ$ to 180$^\circ$, and which 
is normalized according to Eq.~(\ref{eq_normalization}).
Figure~\ref{fig_fitmeasurement} shows the calculated synthetic scattering 
matrix elements. 
The elements of the synthetic scattering matrix are also listed 
in Table~\ref{table1} at an angular resolution of 1 to 5 degrees.
An electronic table of the synthetic scattering matrix elements 
will be available from the Amsterdam 
Light Scattering Database\footnotemark[\value{footnote}] 
\citep[][]{2005JQSRT..90..191V,2006JQSRT.100..437V}.
As a check we used the synthetic scattering matrix to compute 
the same ratios of elements as have been measured. We found the differences
to lie within the ranges of experimental uncertanties or very nearly so.
 
\section{Summary and discussion}
\label{section_summary}

We present measured ratios of elements of the scattering matrix of 
irregularly shaped Martian analogue palagonite particles 
\citep[][]{1995JGR...100.5309R,1997JGR...10213341B} as functions of the scattering angle $\Theta$ 
($3^\circ \leq \Theta \leq 174^\circ$) at a wavelength of 632.8~nm.
Our measured ratios of scattering matrix elements differ strongly 
from those calculated for homogeneous, spherical particles with 
the same size and refractive index.
In particular, the measured phase function 
(ratio $F_{11}(\Theta)/F_{11}(30^\circ)$) 
shows a very strong (almost three orders of magnitude) forward scattering 
peak and a smooth drop-off towards the largest scattering angles, where the 
phase function of the spherical particles shows much more angular features 
especially in the backward scattering direction. Clearly, using scattering matrix
elements that have been calculated for homogeneous, spherical particles when irregularly 
shaped particles are to be expected in radiative transfer calculations for e.g. 
the interpretation of remote-sensing observations can thus lead to errors in retrieved
dust properties (microphysical parameters and/or optical thicknesses). 

To facilitate the use of our measurements in radiative transfer calculations for 
e.g. Mars, we have first constructed an auxiliary scattering matrix from 
the measured scattering matrix. This auxiliary scattering matrix covers 
the whole scattering angle range (i.e. from 0$^\circ$ to 180$^\circ$), 
and its elements have been normalized such that 
the average of the phase function over all scattering directions equals unity. 
The value of the phase function at $\Theta=0^\circ$ has been computed from the phase 
function calculated with Mie-theory for homogeneous, spherical particles with the same size 
and composition. The normalization of the auxiliary phase function, and hence the 
normalization of the other elements too, is obtained by applying a Singular 
Value Decomposition (SVD) method to fit an expansion in generalized spherical 
functions \citep[][]{1963Gelfand,2004Hovenier}  
to the measured phase function, including artificial data points. 
The first expansion coefficient yields the required 
normalization constant. After the normalization of the auxiliary phase function, 
the SVD method is applied to the auxiliary scattering matrix
and its expansion coefficients are obtained. 

With the expansion coefficients, a synthetic scattering matrix is computed 
for the complete scattering range. It is normalized so that the average 
of its one-one element over all directions equals unity.
The synthetic scattering matrix elements can also straightforwardly be used in 
radiative transfer calculations. The need to include all scattering matrix elements,
instead of only the phase function, is obvious for the interpretation 
of polarization observations. However, even for flux calculations, all scattering 
matrix elements should be used, because ignoring polarization, i.e. using only 
the phase function, in multiple scattering calculations induces errors in calculated
fluxes \citep[e.g.][]{1998GeoRL..25..135L}. The use of only the phase function should 
be limited to single scattering calculations for unpolarized incident light. 

\begin{figure}
\begin{center}
\vspace*{-5cm}
\hspace*{-2cm}
\includegraphics[width=16.0cm]{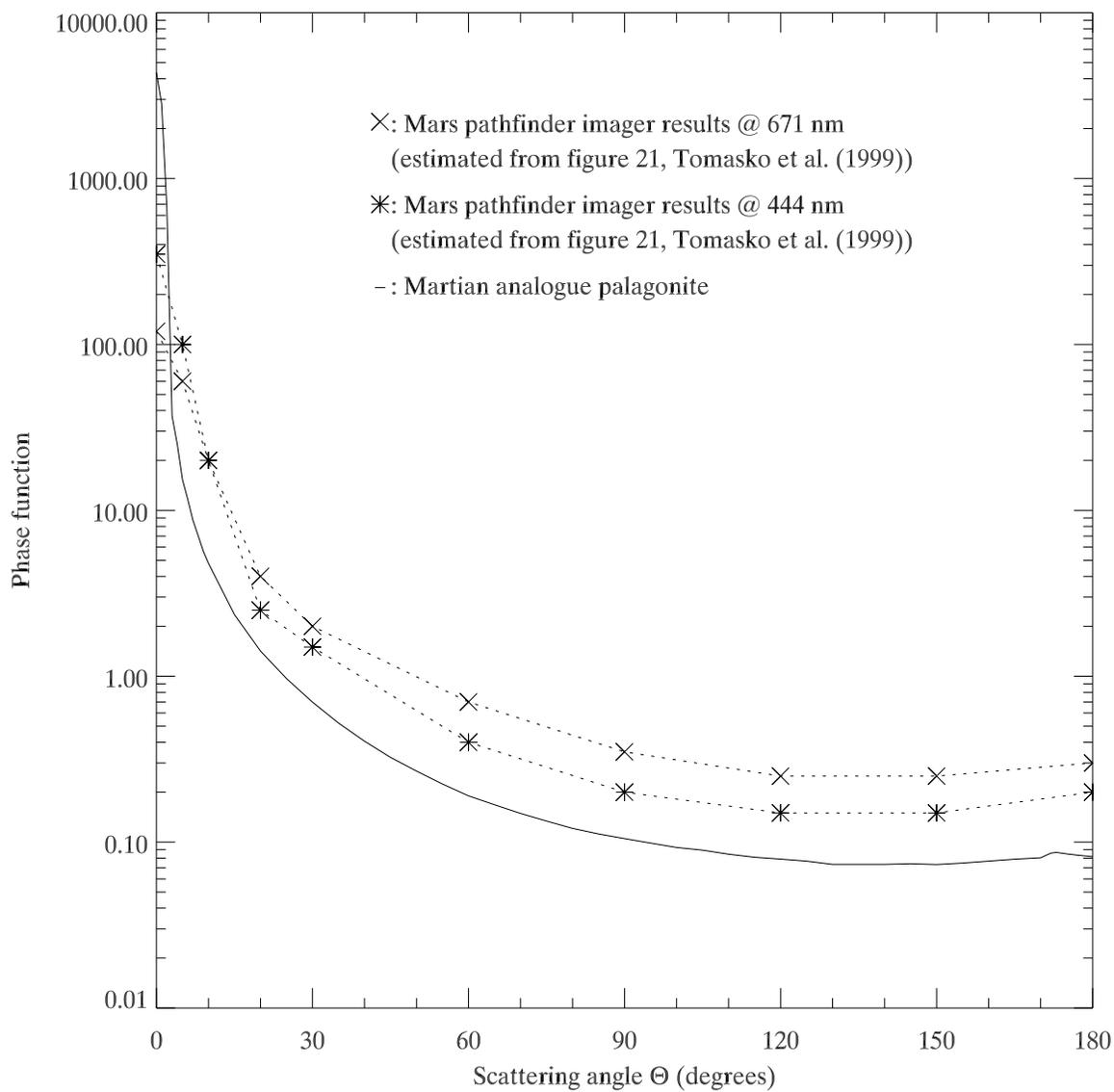}
\vspace*{-0cm}
\end{center}
\caption{The synthetic phase function of the Martian analogue palagonite 
         particles compared to the phase function derived by \citet{1999JGR...104.8987T} from Mars 
         Pathfinder results. Each of the curves has its own normalization.}
\label{fig_map_vs_tomasko}
\label{lastfig}
\end{figure}

Figure~\ref{fig_map_vs_tomasko} shows a comparison between our synthetic phase function 
and two phase functions presented by \citet{1999JGR...104.8987T}
as derived from diffuse skylight observations of the Imager for Mars Pathfinder. Each of the 
phase functions has its own normalization. The phase functions of \citet{1999JGR...104.8987T} show
the same general angular behavior as our synthetic phase function: a strong forward 
scattering peak and a smooth drop-off towards the largest scattering angles. 
The forward scattering peak of our phase function appears to be stronger than 
the peaks of the phase functions of \citet{1999JGR...104.8987T}. This can easily be due to 
the size difference of the particles. The dust particles in our sample have an 
effective radius that is a factor of 2 to 3 larger than that of \citet{1999JGR...104.8987T}
(4.46~$\mu$m versus 1.6~$\mu$m $\pm$ 0.15~$\mu$m). The slopes of the phase functions in 
the backward scattering direction are very similar, and appear to be typical 
for (terrestrial) irregularly shaped mineral particles with moderate refractive 
indices \citep[see e.g.][]{2001JGR...10617375V,2000A&A...360..777M,2001JGR...10622833M}. 
This smooth slope can not be easily anticipated from Mie-theory. 

The expansion coefficients, the synthetic scattering matrix elements, 
and the measured ratios of elements of the scattering matrix of the Martian 
analogue palagonite particles will all be available from the Amsterdam Light 
Scattering Database
\citep[for details, see][]{2005JQSRT..90..191V,2006JQSRT.100..437V}.
The Amsterdam Light Scattering Database contains a collection of measured 
scattering matrix element ratios, including information on particle sizes and 
their composition, for various types of irregularly shaped particles. The SVD 
method presented in this article can straightforwardly be applied to measured 
scattering matrix elements of particles other than the Martian analogue palagonite
particles. 

\begin{figure}
\vspace*{5.0cm}
\begin{center}
\includegraphics[width=12.0cm]{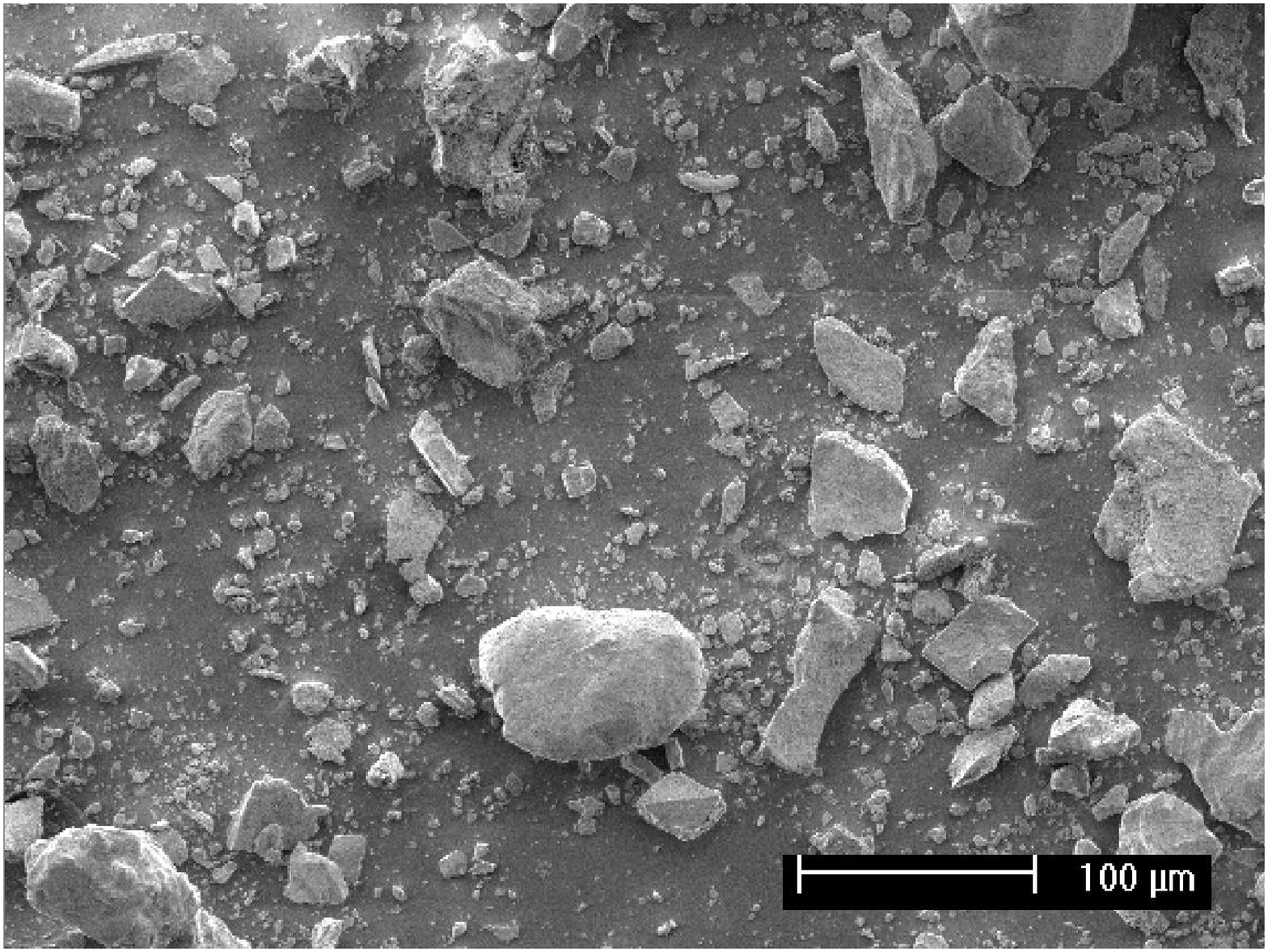}
\end{center}
\vspace*{1.0cm}
\caption{Scanning electron microscope (SEM) image of the Martian 
	analogue palagonite particles used in this study. 
	Note that SEM images generally give a good indication of 
	the typical shapes of particles, but not necessarily of 
	their sizes.}
\label{fig_sem}
\end{figure}

\ack
We are grateful to Ben Veihelmann for helping with the SEM
image of Fig.~\ref{fig_sem}. 
It is a pleasure to thank Martin Konert of the Vrije Universiteit in
Amsterdam for performing the size distribution measurements and Michiel Min
of the University of Amsterdam for fruitful discussions.

\label{lastpage}

                                                                                 
\bibliography{refs_laan2008}
                                                                                 
\bibliographystyle{elsart-harv}

\clearpage
\newpage

\clearpage
\newpage

%
\clearpage
\newpage

\clearpage
\newpage

\clearpage
\newpage

\clearpage
\newpage

\clearpage
\newpage

\clearpage
\newpage

\renewcommand{\arraystretch}{1.2}
 
\newcommand{\cca}{\hspace*{0.3cm} $\Theta$ \hspace*{0.3cm}}
\newcommand{\ccb}{\hspace*{0.2cm} $a_1(\Theta)$ \hspace*{0.2cm}}
\newcommand{\ccc}{\hspace*{0.2cm} $a_2(\Theta)$ \hspace*{0.2cm}}
\newcommand{\ccd}{\hspace*{0.2cm} $a_3(\Theta)$ \hspace*{0.2cm}}
\newcommand{\cce}{\hspace*{0.2cm} $a_4(\Theta)$ \hspace*{0.2cm}}
\newcommand{\ccf}{\hspace*{0.2cm} $b_1(\Theta)$ \hspace*{0.2cm}}
\newcommand{\ccg}{\hspace*{0.2cm} $b_2(\Theta)$ \hspace*{0.2cm}}
\newcommand{\emp}{\hspace*{0.0cm} $ $ }

\newcommand{\xy}{\hspace*{0.07cm}}
\newcommand{\xz}{\hspace*{0.26cm}}
\newcommand{\xw}{\hspace*{0.46cm}}

\begin{table}[h]
\begin{tabular}{|r|r|r|r|r|r|r|} \hline
\cca & \ccb & \ccc & \ccd & \cce & \ccf & \ccg \\ \hline
  0 &  4.36E+003 &  4.26E+003 &  4.26E+003 &  4.12E+003 &  0.00E+000 &  0.00E+000 \\
  1 &  2.82E+003 &  2.75E+003 &  2.75E+003 &  2.66E+003 & -2.34E-003 & -7.15E-003 \\
  2 &  6.60E+002 &  6.43E+002 &  6.43E+002 &  6.17E+002 &  1.98E-002 & -2.68E-002 \\
  3 &  3.71E+001 &  3.58E+001 &  3.62E+001 &  3.49E+001 &  7.32E-002 & -5.39E-002 \\
  4 &  2.51E+001 &  2.43E+001 &  2.46E+001 &  2.38E+001 &  9.69E-002 & -8.20E-002 \\
  5 &  1.53E+001 &  1.46E+001 &  1.49E+001 &  1.43E+001 &  7.28E-002 & -1.05E-001 \\
  6 &  1.17E+001 &  1.13E+001 &  1.14E+001 &  1.10E+001 &  4.88E-002 & -1.18E-001 \\
  7 &  8.75E+000 &  8.44E+000 &  8.49E+000 &  8.18E+000 &  3.73E-002 & -1.20E-001 \\
  8 &  7.08E+000 &  6.82E+000 &  6.82E+000 &  6.60E+000 &  1.66E-002 & -1.12E-001 \\
  9 &  5.68E+000 &  5.48E+000 &  5.48E+000 &  5.29E+000 & -1.68E-003 & -9.83E-002 \\
 10 &  4.80E+000 &  4.62E+000 &  4.61E+000 &  4.42E+000 & -1.55E-003 & -8.22E-002 \\
 15 &  2.35E+000 &  2.22E+000 &  2.21E+000 &  2.12E+000 & -4.94E-003 & -4.04E-002 \\
 20 &  1.42E+000 &  1.32E+000 &  1.31E+000 &  1.23E+000 & -8.44E-003 & -2.15E-002 \\
 25 &  9.68E-001 &  8.93E-001 &  8.80E-001 &  8.13E-001 & -9.63E-003 & -1.35E-002 \\
 30 &  6.98E-001 &  6.27E-001 &  6.18E-001 &  5.60E-001 & -1.05E-002 & -3.85E-003 \\
 35 &  5.22E-001 &  4.57E-001 &  4.47E-001 &  4.08E-001 & -1.15E-002 &  2.72E-003 \\
 40 &  4.06E-001 &  3.44E-001 &  3.32E-001 &  3.00E-001 & -1.30E-002 &  8.12E-003 \\
 45 &  3.24E-001 &  2.68E-001 &  2.50E-001 &  2.25E-001 & -1.46E-002 &  9.34E-003 \\
 50 &  2.68E-001 &  2.13E-001 &  1.96E-001 &  1.75E-001 & -1.55E-002 &  1.23E-002 \\
 55 &  2.24E-001 &  1.71E-001 &  1.52E-001 &  1.38E-001 & -1.66E-002 &  1.46E-002 \\
 60 &  1.90E-001 &  1.37E-001 &  1.20E-001 &  1.10E-001 & -1.67E-002 &  1.57E-002 \\
 65 &  1.68E-001 &  1.15E-001 &  9.64E-002 &  8.91E-002 & -1.76E-002 &  1.59E-002 \\
 70 &  1.49E-001 &  9.70E-002 &  7.72E-002 &  7.20E-002 & -1.74E-002 &  1.80E-002 \\
 75 &  1.34E-001 &  8.14E-002 &  5.95E-002 &  5.79E-002 & -1.71E-002 &  1.79E-002 \\
 80 &  1.21E-001 &  6.99E-002 &  4.60E-002 &  4.70E-002 & -1.68E-002 &  1.82E-002 \\
 85 &  1.12E-001 &  6.03E-002 &  3.58E-002 &  3.79E-002 & -1.62E-002 &  1.85E-002 \\
 90 &  1.05E-001 &  5.30E-002 &  2.48E-002 &  3.12E-002 & -1.66E-002 &  1.77E-002 \\
 95 &  9.87E-002 &  4.73E-002 &  1.86E-002 &  2.50E-002 & -1.52E-002 &  1.71E-002 \\
100 &  9.29E-002 &  4.19E-002 &  1.16E-002 &  2.00E-002 & -1.47E-002 &  1.55E-002 \\
105 &  8.94E-002 &  3.81E-002 &  4.11E-003 &  1.57E-002 & -1.34E-002 &  1.40E-002 \\
110 &  8.46E-002 &  3.42E-002 & -9.87E-004 &  1.16E-002 & -1.20E-002 &  1.35E-002 \\
\hline
\end{tabular}
\vspace*{0.5cm}
\caption{The synthetic scattering matrix elements of the Martian 
         analogue palagonite particles 
         \citep[][]{1995JGR...100.5309R}, as functions of the scattering angle $\Theta$ (in degrees). 
         The scattering angles are identical to those used in the 
         measurements, extended with $\Theta=0^\circ, 1^\circ, 2^\circ$ and from $175^\circ$ to 
         $180^\circ$ degrees.
         An electronic version of this table will be available from the Amsterdam
         Light Scattering Database \citep[][]{2005JQSRT..90..191V},
         at \mbox{http://www.astro.uva.nl/scatter}.}
\label{table1}
\end{table}
 
\addtocounter{table}{-1}
 
\begin{table}[h]
\begin{tabular}{|r|r|r|r|r|r|r|} \hline
\cca & \ccb & \ccc & \ccd & \cce & \ccf & \ccg \\ \hline
115 &  8.10E-002 &  3.25E-002 & -6.60E-003 &  7.86E-003 & -1.19E-002 &  1.22E-002 \\
120 &  7.90E-002 &  3.06E-002 & -1.12E-002 &  4.75E-003 & -1.07E-002 &  1.05E-002 \\
125 &  7.68E-002 &  2.86E-002 & -1.46E-002 &  1.22E-003 & -8.74E-003 &  9.90E-003 \\
130 &  7.34E-002 &  2.75E-002 & -1.77E-002 & -8.03E-004 & -7.27E-003 &  8.21E-003 \\
135 &  7.34E-002 &  2.77E-002 & -2.09E-002 & -3.23E-003 & -6.32E-003 &  6.43E-003 \\
140 &  7.34E-002 &  2.81E-002 & -2.39E-002 & -5.64E-003 & -4.64E-003 &  5.94E-003 \\
145 &  7.41E-002 &  2.87E-002 & -2.62E-002 & -7.85E-003 & -3.46E-003 &  5.51E-003 \\
150 &  7.33E-002 &  2.97E-002 & -2.80E-002 & -9.76E-003 & -2.55E-003 &  3.90E-003 \\
155 &  7.48E-002 &  3.13E-002 & -3.01E-002 & -1.26E-002 & -8.35E-004 &  3.86E-003 \\
160 &  7.68E-002 &  3.47E-002 & -3.50E-002 & -1.38E-002 &  5.71E-005 &  1.69E-003 \\
165 &  7.89E-002 &  3.62E-002 & -3.79E-002 & -1.58E-002 &  1.09E-003 & -9.02E-005 \\
170 &  8.04E-002 &  3.77E-002 & -3.95E-002 & -1.74E-002 &  2.37E-003 & -1.02E-003 \\
171 &  8.32E-002 &  3.90E-002 & -4.17E-002 & -1.82E-002 &  1.96E-003 & -1.26E-003 \\
172 &  8.59E-002 &  4.07E-002 & -4.43E-002 & -1.81E-002 &  1.65E-003 & -1.73E-003 \\
173 &  8.67E-002 &  4.12E-002 & -4.40E-002 & -1.87E-002 &  1.82E-003 & -2.31E-003 \\
174 &  8.59E-002 &  4.10E-002 & -4.65E-002 & -1.62E-002 &  2.12E-003 & -2.78E-003 \\
175 &  8.50E-002 &  4.16E-002 & -4.87E-002 & -1.38E-002 &  2.42E-003 & -2.92E-003 \\
176 &  8.42E-002 &  4.25E-002 & -5.02E-002 & -1.31E-002 &  2.31E-003 & -2.60E-003 \\
177 &  8.35E-002 &  4.42E-002 & -5.07E-002 & -1.36E-002 &  1.85E-003 & -1.88E-003 \\
178 &  8.29E-002 &  4.59E-002 & -5.11E-002 & -1.49E-002 &  1.39E-003 & -9.97E-004 \\
179 &  8.23E-002 &  4.90E-002 & -5.00E-002 & -1.70E-002 &  5.81E-004 & -2.76E-004 \\
180 &  8.18E-002 &  5.04E-002 & -5.04E-002 & -1.89E-002 &  0.00E+000 &  0.00E+000 \\
\hline

\end{tabular}
\vspace*{0.5cm}
\caption{Continued.}
\end{table}

\end{document}